\RequirePackage{fix-cm}
\documentclass[smallextended]{svjour3}       
\smartqed  
\usepackage{graphicx}
\usepackage[dvipsnames,svgnames,x11names]{xcolor}

\usepackage{booktabs}
\usepackage{graphicx}
\usepackage{dcolumn}
\usepackage{bm}
\usepackage{array}

\usepackage{amsmath}
\usepackage{amsfonts}
\usepackage{amssymb}
\usepackage{mathtools}

\usepackage{tikz}

\usepackage{amsbsy}

\newcommand{\ra}[1]{\renewcommand{\arraystretch}{#1}}

\newcommand{\E}{\mathrm{E}}
\newcommand{\dx}[1]{\ \text{d} #1}

\newcommand{\jxadd}{}
\newcommand{\mary}{}
\newcommand{\jx}{}

\begin{document}


\title{Computational tools for assessing gene therapy under branching process models of mutation}



\author{Timothy C. Stutz         \and
        Janet S. Sinsheimer \and 
        Mary Sehl \and 
        Jason Xu*
}


\institute{Timothy C. Stutz \at
              Department of Computational Medicine \\ University of California, Los Angeles, CA \\
              \email{stutztim@ucla.edu}             
           \and
          Janet S. Sinsheimer \at
              Departments of Biostatistics, Computational Medicine, Human Genetics \\
              University of California, Los Angeles, CA \\
              \email{jsinshei@g.ucla.edu}
                \and
            Mary E. Sehl \at 
            Department of Computational Medicine and Division of Hematology-Oncology, Department of Medicine \\ 
            David Geffen School of Medicine, University of California, Los Angeles, CA \\
            \email{msehl@mednet.ucla.edu} 
              \and 
             Jason Xu \at 
            Department of Statistical Science \\
            Duke University, Durham, NC \\
             \email{jason.q.xu@duke.edu}\\
*corresponding author}

\date{Received: date / Accepted: date}

\maketitle

\begin{abstract}
Multitype branching processes are ideal for studying the population dynamics of stem cell populations undergoing mutation accumulation over the years following transplant. In such stochastic models, several quantities are of clinical interest as insertional mutagenesis carries the potential threat of leukemogenesis following gene therapy with autologous stem cell transplantation.  In this paper, we develop a three-type branching process model describing accumulations of mutations in a population of stem cells distinguished by their ability for long-term self-renewal.  Our outcome of interest is the appearance of a double-mutant cell, which carries a high potential for leukemic transformation. In our model, a single-hit mutation carries a slight proliferative advantage over a wild-type stem cells.  We compute marginalized transition probabilities that allow us to capture important quantitative aspects of our model, including  the probability of observing a double-hit mutant and relevant moments of a single-hit mutation population over time.  We thoroughly explore the model behavior numerically, varying birth rates across the initial sizes and populations of wild type stem cells and single-hit mutants, and compare the probability of observing a double-hit mutant under these conditions.  We find that increasing the number of single-mutants over wild-type particles initially present has a large effect on the occurrence of a double-mutant, and that it is relatively safe for single-mutants to be quite proliferative, provided the lentiviral gene addition avoids creating single mutants in the original insertion process.  Our approach is broadly applicable to an important set of questions in cancer modeling and other population processes involving multiple stages, compartments, or types.   
\end{abstract}

\section{Introduction}
\label{intro}

Branching processes comprise a class of stochastic processes that model how a collection of discrete individuals  changes over time, typically by reproducing, dying, or transforming type. 
These processes were first studied in relation to the extinction of family names among the aristocracy by Francis Galton and the Reverend Henry Watson \cite{kendall1966branching}.   
Branching processes have since become prevalent across a wide array of scientific disciplines, from phylogenetics \cite{blum2006random} to cancer biology \cite{durrett2015branching} and nuclear physics \cite{pazsit2007neutron}.  In systems biology contexts related to our interest here, branching processes have been extensively applied to model clonal expansion in response to tumor suppressor gene inactivation  \cite{michor2006stochastic,nowak2006genetic} and the evolutionary dynamics of leukemia in response to combination therapy
\cite{komarova2005drug,wodarz2009towards}.  \mary{Stochastic models can be used to study events such as  extinction of a leukemic clone or mutation leading to malignant transformation.}

While many properties of branching processes are well-studied, computing finite-time quantities that characterize their transient behavior remains challenging. 
In this article we propose numerical methods for computing a wide range of quantities pertinent to a multi-type branching process system evolving in continuous time.
We are motivated by the biological process of mutation accumulation, specifically in clonal hematopoiesis leading to myelodysplasia and leukemia, and in gene therapy through lentiviral gene addition. 
To this end, we consider a three-type branching process model for stagewise mutation in a reproducing population of stem cells, and show how a variety of marginalized transition probabilities that are useful toward addressing biological questions of interest. 
This method also enables us to compute the expected sizes of sub-populations at any time $t$ and related probabilistic quantities with ease in a general $m$-type model.

A classical mathematical tool used for studying branching processes is the probability generating function (PGF).
The PGF greatly simplifies the computation of transition probabilities, the fundamental quantities that determine how a stochastic process behaves.
One avenue of attack is to work with an infinite series of coupled ordinary differential equations (ODEs), frequently called the Master Equations, that govern the evolution of the \jxadd{processs states}. 
Instead we exploit the independence of particle lineages to reduce the computation to solving a small number of ODEs \jxadd{governing the behavior of the PGFs for processes} originating with a single particle of each type.  
\jxadd{This property of branching processes---that particles act independently according to a probability law---makes these simplified mathematical manipulations possible,} greatly simplifying their computational cost.

Our \jxadd{method} 
revolves around three core steps.
We first factorize the PGF for the full process into a product of PGFs of the process beginning with a single particle of each type.
Second, we solve for these PGFs using the pseudo-generating function technique which involves a simple numerical computation of a small series of coupled ODEs \cite{bailey1990elements} governing the evolution of the PGF. 
This is followed by fast numerical inversion based on the Fast Fourier transform to recover the desired quantities from the PGF solutions.

Much of the prior mathematical literature has largely focused on the ``forward" behavior of branching processes,  including the long-run or limiting behavior of \jxadd{such processes} \cite{durrett2015branching,bozic2013evolutionary}. 
The derivations presented in this article instead \jx{help us quantify transient behavior with an efficient and general numerical recipe, extending these methods beyond the special cases for which solutions have been previously established}. In particular, finite-time quantities often translate directly to answer relevant clinical questions, and are necessary within inferential schemes \cite{xu2015likelihood,xu2019statistical} such as evaluating the marginal likelihood of data available at a sequence of observation times. Indeed, transition probabilities fully characterize the probabilistic behavior of these stochastic models. Our approach targets these and related quantities applicable to  multi-type branching processes beyond our model of mutation here. This contrasts with previous work 
that for instance posits a neutral or symmetric branching process with specific relationships between the birth and death parameters \cite{lan2017fate}; it is worth noting that the more general version of that model was deemed mathematically intractable, analyzed only in supplementary simulations.    
Our contributions fill a methodological gap suggested in this and related prior literature, which largely resorts to brute-force simulation and thus requires considerable computation time \cite{lan2017fate}. 

\mary{As a motivating application of these methods, we consider a numerical examination of gene therapy in hematopoietic stem cell lines. Advances in gene therapy offer a promising approach to the treatment of genetic hematologic diseases and immunodeficiency disorders \cite{kohn2001gene,cavazzana2000gene,hacein2002sustained}.  Lentiviral vectors, derived from the human immunodeficiency virus, can efficiently insert, modify, and delete genes into a cell's genome. They have been extensively investigated and optimized over the last twenty years, and their use is commonplace in academic laboratories and in gene therapy clinical trials, such as in the treatment of primary immunodeficiencies and hematologic disorders \cite{reinhardt2021long}. Most recently, gene editing through CRISPR technology is being investigated as a means to treat inherited diseases and offers hope for future translational work.  However, for any gene therapy, when a corrected gene is being inserted into the genome, it carries a chance of being inserted in a site that is close to an oncogene.  As a result, induction of leukemia remains a major concern for gene therapy clinical trials \cite{kohn2001gene,marshall2002gene,marshall2003second}, although more recently developed methods adopt risk reduction strategies to mitigate these concerns  \cite{reinhardt2021long}.  Because insertion of a corrected gene can disrupt gene structure and alter gene transcription and gene regulation, there is always a theoretical risk of cell transformation, although reported events are extremely rare. In the historical setting of retroviral gene therapy studies, these exceedingly rare events may occur within weeks of treatment or even up to years later \cite{kohn2003occurrence}.  Further risk of mutational events in patients undergoing gene therapy clinical trials comes from the chemotherapy regimens which carry a risk of hematopoietic stem cell damage and leukemic transformation.

Because leukemia is an unacceptable outcome in any clinical trial, it is important to quantify the risk of its development in this setting} \cite{bonetta2002leukemia}.  Our goal is to quantify the likelihood of leukemic transformation and identify conditions under which a leukemic clone is likely to expand.
Using our numerical methods we explore the population dynamics of transduced hematopoietic stem cells after stem cell transplantation under a multitype branching process model. This stochastic modeling approach can quantify the probability of observing rare events under varying clinical scenarios, such as mutational oncogenesis and engraftment failure that may occur during stem cell transplant.

We will investigate \jxadd{multiple questions clinically relevant to gene therapy here}, including \mary{1)} the time until two \mary{mutational} events occur leading to leukemogenesis, and \mary{2) } how varying the proliferative advantage of the transduced population alters the probability of observing leukemic outcomes. These efforts are motivated by Knudson's finding that two rate-limiting mutational steps are required for inactivation of tumor suppressor genes in sporadic cancers \cite{knudsen1971mutation}, and apply generally to work studying initiation and progression of tumors and identifying driver mutations. 
We conclude by outlining how our methods may be applied to other clinical scenarios in cancer research such as the inactivation of the two alleles of a tumor suppressor gene and the acquisition of multiple driver mutations.

\section{Mathematical Notation for Branching Processes}
\label{notation}

We begin with a multi-type branching process whose counts are denoted by the vector $\mathbf{X}(t)$.  A continuous-time branching process $\mathbf{X}(t)$ is a Markov jump process in which a collection of individuals, in our case cells, can reproduce and die independently according to a probability distribution. Each component $X_i(t)$ denotes the number of particles of type $i$ at time $t \geq 0 $. Each type has a distinct mean lifespan and reproductive probabilities, and can give rise to cells of its own type as well as other types at its time of death.

\begin{figure}[ht]
\begin{center}
\begin{tikzpicture}[
roundnode/.style={circle, draw = black, thick, minimum size=13mm},
]
\node[roundnode] (x1) at (0cm, 0cm)	{$X_1$} ;
\node[roundnode] (x2) at (3cm, 0cm) {$X_2$} ;

\draw[->] ([xshift = +0.06cm, yshift = -0.06cm]x1.315) .. controls +(0.3, -1.25) and +(-0.3, -1.25) .. node[below] {$\lambda_1$} ([xshift = -0.06cm, yshift = -0.06cm]x1.225) ;
\draw[->] ([xshift = +0.06cm, yshift = -0.06cm]x2.315) .. controls +(0.3, -1.25) and +(-0.3, -1.25) .. node[below] {$\lambda_2$} ([xshift = -0.06cm, yshift = -0.06cm]x2.225) ;

\draw[->] ([xshift = +0.06cm, yshift = +0.06cm]x1.65) to node[above left] {$\mu_1$} +(0.7cm, 0.7cm) node[xshift = +0.2cm, yshift = +0.2cm] {$\emptyset$} ; 
\draw[->] ([xshift = +0.06cm, yshift = +0.06cm]x2.65) to node[above left] {$\mu_2$} +(0.7cm, 0.7cm) node[xshift = +0.2cm, yshift = +0.2cm] {$\emptyset$} ; 

\draw[->] ([xshift = +0.1cm]x1.0) to node[above] {$\nu$} ([xshift = -0.1cm]x2.180)  ;
\end{tikzpicture}
\end{center}
\caption{Reaction diagram for the two-type branching process.  $\lambda_i$ corresponds to reproduction through binary fission, $\mu_i$ corresponds to removal via cell death, and $\nu$ corresponds to transformation via mutation.}
\label{fig:4two}
\end{figure}
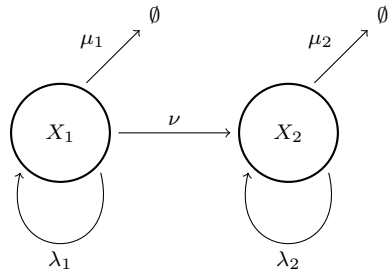
\jx{To simplify the exposition and notational load, we focus our attention on the two-type case depicted in Figure \ref{fig:4two} as we present the definitions and derivations. We then clarify how the methodology applies in the general case, including the three-type method we implement in our numerical analysis of mutation in hematopoietic stem cells.}
Let $a_i(k,l)$ denote the rate at which a particle of type $i$ produces $k$ particles of type one and $l$ particles of type two upon completion of its lifespan. 
For example, $a_1(2,0)$ is the rate at which a single particle of type one reproduces, producing two offspring through binary fission.
We define the negative total rate as
\[ \alpha_i \equiv a_i(1,0) = - \sum_{(k,l) \neq (1,0)} a_i(k,l)     \]
which satisfies a conservation of total rates $\sum_{(k,l)} a_i(k,l) = 0$. 
\jx{Our process is time-homogeneous in that the rates $a_i(k,l)$ do not change over time. This property together with the assumption of particle independence implies rate-linearity:  the reaction rate of any event at time $t$ is given by the per-particle rate multiplied by the corresponding population at time $t$.} 

\jxadd{These observations lead us to the instantaneous jump probabilities that characterize the process. For a short time increment $h$}, jumping from a population of $j$ type one particles to $k$ type one particles and $l$ type two particles is given by
\begin{equation}\label{eq:4expansion}
 \Pr \left( \mathbf{X}(h) = (k,l) \mid \mathbf{X}(0) = (j,0)  \right) = j \cdot a_1(k,l) \cdot h + o(h). 
\end{equation}
 This form \jxadd{is consistent with our branching process being an instance of a time-homogeneous continuous time Markov chain \cite{lange2010applied}.  Note that as $h \downarrow 0$ the instantaneous rate of change is given by $j \cdot a_1(k,l)$ as expected from the previous discusson of rate-linearity. }
By the Markov assumption, the lifespan of each particle of type $i$ is exponentially distributed with rate $-\alpha_i$.  

\jxadd{For an arbitrary interval of time $t$ that need not be close to zero, we denote the finite-time  \textit{transition probabilities}}
\begin{equation}\label{eq:4TPs}
P_{(i,j),(k,l)}(t) \equiv \Pr \left( \mathbf{X}(t) = (k, l) \mid \mathbf{X}(0) = (i, j)  \right)   .
\end{equation}
\jxadd{In contrast to \eqref{eq:4expansion}, the form of $P_{(i,j),(k,l)}(t)$ is typically not readily available, requiring a large integration step accounting for all possible process paths and their respective probabilities passing between the endpoints $\mathbf{X}(0)$ and $\mathbf{X}(t)$ . Towards computing these quantities, we note that they appear within the  probability generating function (PGF), given by}
\begin{equation}\label{eq:4PGF}
 \phi_{ij}(t, s_1, s_2) \equiv \sum_{k=0}^\infty \sum_{j=0}^\infty P_{(i,j),(k,l)}(t) s_1^k s_2^l   = \E \left[ s_1^{X_1(t)} s_2^{X_2(t)} \mid X_1(0) = i, X_2(0) = j  \right]
 \end{equation}
where the dummy variables $s_i \in \mathbb{C}$ satisfy $\| s_i \| \leq 1$ .  
We will frequently shorten the notation for the PGF to $\phi_{ij}(t)$, suppressing dependence on the dummy variables.

The PGF is a primary mathematical object playing a key role in our computational technique.
Notably, given numerical methods to compute the PGF, it is \jxadd{tractable to invert the PGF using the Fast Fourier Transform to obtain  transition probabilities \eqref{eq:4TPs}, discussed further below.}
This is not an entirely trivial numerical task, however, requiring $O \left(  \max (k,l)^2\right) $ computations \cite{xu2015efficient}.  

Fortunately, it is often the case that we are concerned primarily with the distribution of a single sub-population. \jxadd{For instance, in assessing the safety of a gene therapy we may typically pose questions pertaining to the number of particles in the mutant or type two population.} 
In such cases, the relevant series expression can be obtained by \jxadd{\textit{marginalizing} out all other types from the joint distribution of the process.}
Fortunately, doing so becomes straightforward \jxadd{if one has the joint PGF in hand and further provides the advantage of reducing the computational complexity considerably}. Marginalizing out the first type, for instance, is accomplished by letting $s_1 = 1$ in the PGF \eqref{eq:4PGF},
\[ \phi_{ij}(t, s_1, s_2) \big|_{s_1 = 1} = \sum_{l = 0}^\infty P_{(i,j),(\cdot,l)}(t) s_2^l .  \] 
Obtaining these marginalized probabilities is much more computationally efficient, as it requires inverting only a univariate probability generating function \jxadd{\textit{regardless} of the number of total types in the branching process system.  } \jx{Indeed, though the notation above corresponds to the two-type case, the analogous quantity pertaining to a type $j$ population in the general multitype case can be computed by setting $s_i=1$ for all populations $i\neq j$. }


\section{Computing the Probability Generating Function}
\label{PGF}
\jx{Here we detail how to compute the probability generating function following a classical technique detailed in \cite{bailey1990elements}.} We shall generate a series of Ordinary Differential Equations (ODEs) that govern the evolution of the PGF over time.  
We first make the observation that, by particle independence,
\begin{equation}\label{eq:4PGFIndep}
\phi_{ij}(t) = \phi_{(1,0)}(t)^i \phi_{(0,1)}(t)^j .
\end{equation}
\jx{For simplicity's sake we shall shorten the notation $\phi_{(1,0)} \equiv \phi_1$ and $\phi_{(0,1)} \equiv \phi_2$}. Since each clan of particles is independent of every other clan, we can view the full PGF as being composed of the individual contributions from each original ancestor particle present at time $t = 0$. \jx{We emphasize here that although the branching process is defined over a countably infinite state space, by making use of the master equations applied to the PGF instead of the process itself, we need only to consider a system of \textit{two} differential equations. Moreover, this technique is general and entails $m$ differential equations for the $m$-type case, which we discuss after detailing the two-type case in order to lighten notation. }

%

Equation \eqref{eq:4PGFIndep} suggests it suffices to work with the PGFs of the process starting with a particle of each given type. Again illustrating with two-type notation, consider the small-time expansion of $\phi_1$ using \eqref{eq:4expansion},
\begin{align}\label{eq:4phi1}
\phi_1& (t, s_1, s_2) = \E \left[ s_1^{X_1(t)} s_2^{X_2(t)} \mid X_1(0) = 1, X_2(0) = 0  \right] = \sum_{k=0}^\infty \sum_{l=0}^\infty  P_{(1,0),(k,l)}(t) s_1^k s_2^l \nonumber \\
& = \sum_{k=0}^\infty \sum_{l=0}^\infty  \left[ \mathbf{1}_{k=1, l=0} + a_1(k,l)t + o(t)   \right] s_1^k s_2^l = s_1  + t \sum_{k=0}^\infty \sum_{l=0}^\infty  a_1(k,l) s_1^k s_2^l + o(t)
\end{align}
as $t \downarrow 0$. Here $\mathbf{1}$ denotes the indicator function. Above, we see that the final expression in Equation \eqref{eq:4phi1} involves the \textit{pseudo-generating function} \cite{bailey1990elements}
\begin{equation}\label{eq:4pseudo}
u_i(s_1, s_2) = \sum_{k=0}^\infty \sum_{l=0}^\infty a_i(k,l) s_1^k s_2^l .
\end{equation}
\jx{Using classical Chapman-Kolmogorov arguments with details in the Appendix, we arrive at the system of backward equations}
\begin{align}
\frac{\dx }{\dx t}\phi_1 (t) &= u_1(\phi_1(t), \phi_2(t)),  \qquad \phi_1(0) = s_1   \nonumber \\
\frac{\dx }{\dx t}\phi_2 (t) &= u_2(\phi_1(t), \phi_2(t)), \qquad \phi_2(0) = s_2 .  \nonumber
\end{align}
The initial conditions follow from the definition of the PGF \eqref{eq:4PGF}.

We will now focus on the specific rates $a_i(k,l)$ that are used in the two-type branching process, listed below:
\begin{eqnarray}
& a_1(2,0) = \lambda_1, 	\qquad	a_1(0,0) = \mu_1, 	\qquad	a_1(0,1) = \nu,	\qquad	a_1(1,0) = - (\lambda_1 + \mu_1 + \nu) \nonumber \\
& a_2(0,2) = \lambda_2, 	\qquad	a_2(0,0) = \mu_2, 	\qquad	a_2(0,1) = - (\lambda_2 + \mu_2 ) . \nonumber 
\end{eqnarray}
Writing these in terms of  pseudo-generating functions, we have
\begin{eqnarray}
&u_1(s_1, s_2) = \lambda_1 s_1^2 +  \nu s_2 - (\lambda_1 + \mu_1 + \nu) s_1 + \mu_1 \nonumber \\
&u_2(s_1, s_2) = \lambda_2 s_2^2 - (\lambda_2 + \mu_2) s_2 + \mu_2 \nonumber .
\end{eqnarray}
We use these expressions in the differential equations for the single-particle generating functions to arrive at the desired backwards equations and initial conditions, as follows:
\begin{align}
\frac{\dx }{\dx t}\phi_1 (t) &= \lambda_1 \phi_1^2 + \nu \phi_2 - (\lambda_1 + \mu_1 + \nu) \phi_1 + \mu_1,  \qquad &\phi_1(0) = s_1   \label{eq:4dphi1} \\
\frac{\dx }{\dx t}\phi_2 (t) &= \lambda_2 \phi_2^2 - (\lambda_2 + \mu_2) \phi_2 + \mu_2, \qquad \qquad &\phi_2(0) = s_2 .  \label{eq:4dphi2}
\end{align}

\subsection{Solving for $\phi_2$}
\label{phi2}

Upon inspection, the differential equation \eqref{eq:4dphi2} is a nonlinear first order Riccati equation, which is fortunately solvable.
This is because the ansatz $\phi_2 = K$ provides us the constant particular solutions $\phi_2 = 1$ and $\phi_2 = \mu_2 / \lambda_2$.  
As is standard for solving such a Riccati equation, we make the substitution $z = \frac{1}{\phi_2 - 1}$, implying that $\phi_2 = 1 + 1/z$.  
Making this substitution in  \eqref{eq:4dphi2} gives
\begin{align*}
\phi_2' = -\frac{z'}{z^2} & = \mu_2 - (\lambda_2 + \mu_2)\left( 1 + \frac{1}{z} \right) + \lambda_2 \left( 1 + \frac{1}{z} \right)^2   \\ 
& = \mu_2 - (\lambda_2 + \mu_2) - \frac{\lambda_2 + \mu_2}{z} + \lambda_2 \left( \frac{1}{z^2} + \frac{2}{z} + 1 \right) \\
& = \frac{\lambda_2 - \mu_2}{z} + \frac{\lambda_2}{z^2} .
\end{align*}
Multiplying through by $-z^2$ and rearranging gives a linear first order differential that is straightforward to solve using an integrating factor,
\begin{align*}
z' & = (\mu_2 - \lambda_2)z - \lambda_2 \\
z  & =  \frac{\lambda_2}{\mu_2 - \lambda_2} + C e^{(\mu_2 - \lambda_2)t}
\end{align*}
for some constant $C$ that depends on the initial conditions.
We substitute $z$ back into $\phi_2 = 1 + 1 / z$ and include the initial conditions $\phi_2(0) = s_2$ to get the solution
\begin{equation}\label{eq:phi2}
\phi_2(t, s_2) = 1 + \left[ \frac{\lambda_2}{\mu_2 - \lambda_2} + \left(\frac{1}{s_2-1} + \frac{\lambda_2}{\lambda_2 - \mu_2} \right)e^{(\mu_2 - \lambda_2)t} \right] ^{-1} .
\end{equation}
Note that in the critical case $\lambda_2 = \mu_2$, Equation \eqref{eq:phi2} becomes
\[ \phi_2(t, s_2) = 1 + \left[ \frac{1}{s_2 -  1} - \lambda_2 t   \right]^{-1} .    \]

\subsection{Solving for $\phi_1$ and the general case}

The differential equation \eqref{eq:4dphi1} governing $\phi_1$ is more difficult due to the presence of a mutation term involving $\phi_2$.  
We now have a single inhomogeneous nonlinear Riccati equation that cannot be solved by a simple substitution trick. 
However, we note that it is straightforward to solve $\phi_1$ numerically using any number of differential equation solvers; we use \verb|DifferentialEquations.jl| package in the \verb|Julia| programming language \cite{rackauckas2017differentialequations,bezanson2017julia}. \jxadd{The computational bottleneck to evaluate the joint PGF at any time $t>0$ amounts to numerically solving a single ODE.}

\begin{figure}[ht]
\begin{center}
\begin{tikzpicture}[
roundnode/.style={circle, draw = black, thick, minimum size=13mm},
]
\node[roundnode] (x1) at (0cm, 0cm)	{$X_1$} ;
\node[roundnode] (x2) at (3cm, 0cm) {$X_2$} ;
\node[roundnode] (x3) at (6cm, 0cm) {$X_3$} ;

\draw[->] ([xshift = +0.06cm, yshift = -0.06cm]x1.315) .. controls +(0.3, -1.25) and +(-0.3, -1.25) .. node[below] {$\lambda_1$} ([xshift = -0.06cm, yshift = -0.06cm]x1.225) ;
\draw[->] ([xshift = +0.06cm, yshift = -0.06cm]x2.315) .. controls +(0.3, -1.25) and +(-0.3, -1.25) .. node[below] {$\lambda_2$} ([xshift = -0.06cm, yshift = -0.06cm]x2.225) ;
\draw[->] ([xshift = +0.06cm, yshift = -0.06cm]x3.315) .. controls +(0.3, -1.25) and +(-0.3, -1.25) .. node[below] {$\lambda_3$} ([xshift = -0.06cm, yshift = -0.06cm]x3.225) ;

\draw[->] ([xshift = +0.06cm, yshift = +0.06cm]x1.65) to node[above left] {$\mu_1$} +(0.7cm, 0.7cm) node[xshift = +0.2cm, yshift = +0.2cm] {$\emptyset$} ; 
\draw[->] ([xshift = +0.06cm, yshift = +0.06cm]x2.65) to node[above left] {$\mu_2$} +(0.7cm, 0.7cm) node[xshift = +0.2cm, yshift = +0.2cm] {$\emptyset$} ; 
\draw[->] ([xshift = +0.06cm, yshift = +0.06cm]x3.65) to node[above left] {$\mu_3$} +(0.7cm, 0.7cm) node[xshift = +0.2cm, yshift = +0.2cm] {$\emptyset$} ; 

\draw[->] ([xshift = +0.1cm]x1.0) to node[above] {$\nu_1$} ([xshift = -0.1cm]x2.180)  ;
\draw[->] ([xshift = +0.1cm]x2.0) to node[above] {$\nu_2$} ([xshift = -0.1cm]x3.180)  ;
\end{tikzpicture}
\end{center}
\caption{Reaction diagram for the three-type branching process.  $\lambda_i$ corresponds to reproduction through binary fission, $\mu_i$ denotes removal via death, and $\nu_i$ represents transformation via mutation.}
\label{fig:4three}
\end{figure}
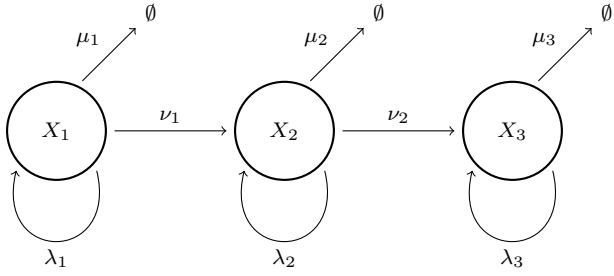

\jx{We have focused on the two-type model to provide a clear exposition to bridge classical methods to the numerical techniques we will utilize, but these immediately extend beyond the two-type case. Given an instance of an $m$-type model in the analogous framework, it is straightforward to write the system of backward equations analogous to \eqref{eq:4dphi1} and \eqref{eq:4dphi2}. Typically, the equation corresponding to the final type may be solvable, while we recommend numerically solving the remaining system of differential equations. For example, the survival probability of a birth-death model has an analytical solution.

Concretely, we illustrate how this approach readily applies to a three-stage model that will be used to study mutation in a hematopoietic model in Section \ref{sec:numerical}.  Figure \ref{fig:4three} illustrates the reaction rates characterizing the model. Mirroring the notation earlier, we abbreviate $\phi_1 = \phi_{(1,0,0)}$, $\phi_2 = \phi_{(0,1,0)}$, and $\phi_3 = \phi_{(0,0,1)}$.  }
The differential equations that govern the single-particle generating functions are thus:
\begin{align*}
\frac{d}{dt} \phi_1 & = \lambda_1 \phi_1^2 - \left( \lambda_1 - \mu_1 - \nu_1 \right) \phi_1 + \nu_1 \phi_2 + \mu_1 	\nonumber \\
\frac{d}{dt} \phi_2 & = \lambda_2 \phi_2^2 - \left( \lambda_2 - \mu_2 - \nu_2 \right) \phi_2 + \nu_2 \phi_3 + \mu_2 	\nonumber \\
\frac{d}{dt} \phi_3 & = \lambda_3 \phi_3^2 - \left( \lambda_3 - \mu_3 \right) \phi_3  + \mu_3 	\nonumber .
\end{align*} 
Note that $\frac{d}{dt}\phi_3$ has the same structure as $\frac{d}{dt}\phi_2$ from the two-type exposition \eqref{eq:4dphi2};  we can use the same argument to show that 
\[ \phi_3(t, s_3) = 1 + \left[ \frac{\lambda_3}{\mu_3 - \lambda_3} + \left(\frac{1}{s_3-1} + \frac{\lambda_3}{\lambda_3 - \mu_3} \right)e^{(\mu_3 - \lambda_3)t} \right] ^{-1} ,  \]
and likewise solve $\phi_1$ and $\phi_2$  numerically. 

\jx{It is worth noting that in special cases such as two-type models described above, it is possible to derive exact solutions  \cite{antal2011exact}. From a numerical perspective, terms such as hypergeometric functions arising in these such solutions can present numerical challenges in some parameter regimes \cite[Appendix~A]{xu2015likelihood}. When this is the case, which often occurs when these solutions are required within iterative schemes, it can be preferable to solve $\phi_1$ using numerical differential equation solvers. Moreover, the numerical recipe we advocate above for the solving the system holds for any $m$-type model in our class.}

\subsection{Inverting the Numerical PGF}

Once we have solved for $\phi_1$ and $\phi_2$ to a specified time point, we can compute the full PGF $\phi_{ij}$ via \eqref{eq:4PGFIndep}.
We then use a technique introduced in \cite{lange1982calculation} to invert the PGF to obtain our desired marginalized univariate probability function.
We begin with a change of variables $s = e^{2 \pi i w}$, placing our PGF argument $s$ on the unit circle in the complex plane.
This transforms the generating function into a periodic function,
\[ \phi_{ij} \left(t, e^{2 \pi i w}  \right)  = \sum_{k = 0}^\infty c_k(t) e^{2 \pi i w  k}  \]
where $c_k(t)$ are the coefficients that contain our desired transition probabilities and $i = \sqrt{-1}$.
$c_k$ is also the $k$th coefficient of a Fourier series, and therefore can be obtained by a straightforward inversion
\[ c_k(t) = \int_0^1 \phi_{ij} \left(t, e^{2 \pi i w}  \right) e^{- 2 \pi i wk}   dw .  \]
In practice, the integral \jxadd{can be evaluated numerically via discretization. A Riemann sum approximation is appropriate for large $N$}, yielding
\[c_k(t) \equiv P_{(ij),k}(t) \approx \frac{1}{N - 1} \sum_{l=0}^{N-1}  \phi_{ij} \left(t, e^{2 \pi i l / N} \right) e^{- 2 \pi i l k / N} . \] 
Note that one can increase the accuracy of the approximation by increasing the size of the gridlength, i.e. make $N$ larger.
Most importantly, rather than naively summing all terms in the Riemann sum, one can apply the Fast Fourier Transform (FFT) to recover \textit{all} of the coefficients for $k = 0 $ to $N-1$ simultaneously in a single computation. 
This yields the marginalized transition probabilities in a numerically stable and efficient manner \jx{regardless of the number of types. In contrast, without using the marginalized forms, this technique would involve summations over each population type, dramatically increasing the computational cost.}

\subsection{Extracting Relevant Probabilities and Expectations}

With a numerical solution for $\phi_{ij}$ now readily available, obtaining expectations becomes straightforward via numerical differentiation. For instance, from starting population $(m,n)$,  one can compute the average type $i$ population after any time interval of length $t$ as
\[ \E \left[ X_i(t) \mid \mathbf{X}(0) = (m, n) \right] = \frac{\partial}{\partial s_i} \phi_{(m,n)}(t, s_1, s_2) \big|_{s_1 = s_2 = 1} .\]
Higher order moments can be obtained in a similar fashion by computing further partial derivatives; see \cite{lange2010applied}. 

While such quantities that are expressible as expectations can also be computed by way of moment generating functions, other more nuanced quantities that translate to specific, interpretable questions often require the full PGF.  Many clinical questions can be stated in terms of thresholds, such as the probability that a double-mutant population exceeds a critical level.
These probabilities can be obtained by summing on the probability generating function of the population of interest after the  threshold, while marginalizing over all other subtypes:
\[ \Pr \left( X_2(t) \geq M \mid \mathbf{X}(0) = (i,j) \right) = \sum_{k = 0}^\infty \sum_{l=M}^\infty  P_{(i,j),(k,l)}(t)  . \]
Similarly, for populations with more than two types, marginalizing over multiple types or species $i$ is accomplished by substituting $s_i = 1$ into the joint PGF.

Indeed, when the question of interest regards a critical level pertaining to one type,  we wish to marginalize out all but one of the populations.
Specifically marginalizing over $X_1$ can be done via
\[ \phi_{ij}(t,s_1,s_2) \big|_{s_1 = 1} = \sum_{k = 0}^\infty \sum_{l=0}^\infty P_{(i,j),(k,l)}(t) s_2^l =  \sum_{l=0}^\infty P_{(i,j),(\cdot,l)}(t)s_2^l.   \]
This marginalized PGF now describes the probabilistic behavior of the $X_2$ population, having correctly accounted for the probabilistic behavior of all configurations of the rest of the system. As such it becomes a univariate function in $s_2$.  The previous numerical FFT approach to invert the generating function becomes computationally efficient to obtain the marginal transition probabilities $P_{(i,j),(\cdot,l)}(t)$ for $X_2(t)$.
Then we compute the desired threshold probabilities as 
\[ \Pr \left( X_2(t) \geq M \mid \mathbf{X}(0) = (i,j)  \right)  =  1 - \sum_{l = 0}^{M-1} P_{(i,j),(\cdot,l)}(t).     \]


\section{Numerical Results}\label{sec:numerical}

%

We now investigate the three-type branching process as a model for diploid mutations in a growing population of hematopoietic stem cells. 
This model can be applied to a growing population of transplanted stem cells in gene therapy, particularly to the acquisition of mutations that lead to oncogenesis.
The rate parameters are listed in Table \ref{table:4parameters}, corresponding to the  reactions depicted previously in Figure \ref{fig:4three}.

\begin{table}[h]
	\ra{1.2}
	\centering
	\begin{tabular}{@{}ccccccc@{}} \toprule
& Rate Symbol 				&& Diagram 	&& Parameter Value	&\phantom{ABCD}  \\ \midrule[0.5pt]
& $\lambda_1$  	&& $X_1  \rightarrow X_1 + X_1 $  	&& $2.4 \times 10^{-2}$    & \\
& $\mu_1$  		&& $X_1  \rightarrow \emptyset $  	&& $1.4 \times 10^{-2}$  	& \\
& $\nu_1$  		&& $X_1  \rightarrow X_2		$  	&& $\lambda_1 \times 10^{-8} $ 	& \\
& $\lambda_2$  	&& $X_2  \rightarrow X_2 + X_2 $  	&& Varies  	& \\
& $\mu_2$  		&& $X_2  \rightarrow \emptyset $  	&& $1.4 \times 10^{-2}$  	& \\
& $\nu_2$  		&& $X_2  \rightarrow X_3		$  	&& $\lambda_2 \times 10^{-8} $  & \\
& $\lambda_3$  	&& $X_3  \rightarrow X_3 + X_3 $  	&& Varies  	& \\
& $\mu_3$  		&& $X_3  \rightarrow \emptyset $  	&& $1.4 \times 10^{-2}$  	& \\
 \bottomrule
\end{tabular}
\caption{Reactions and parameters for the three-type branching process.  The rate parameters correspond to units of 1/week and are sourced from  \cite{abkowitz2002evidence,pillay2010vivo,patel2017fate,michor2005dynamics}.}
\label{table:4parameters}
\end{table}

\begin{figure}[ht]\begin{center}
\includegraphics[width = 0.75\textwidth]{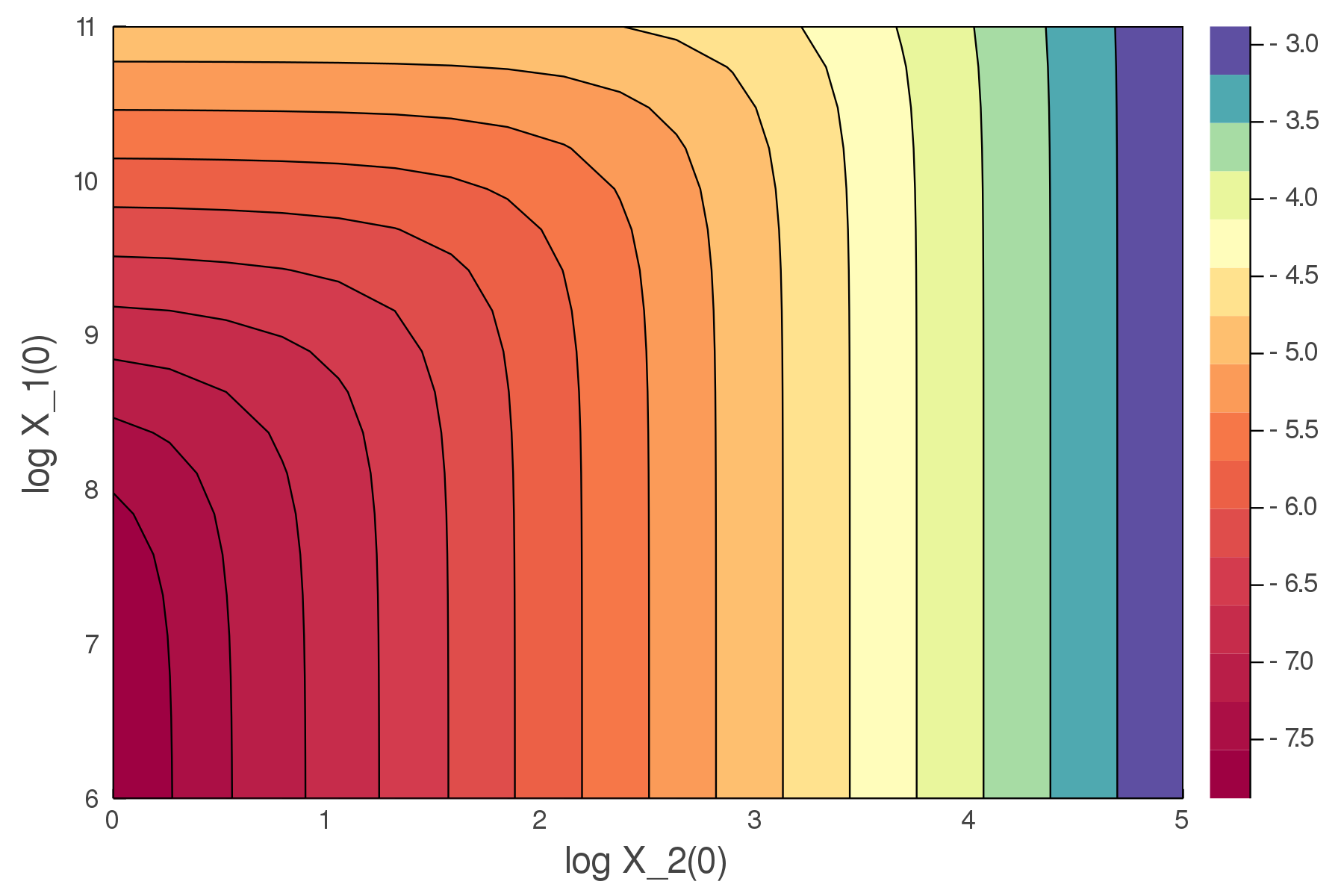}
\end{center}
\caption{Log probability of observing at least one double-mutant at $t = 52$ weeks while varying the initial counts of $X_1$ and $X_2$.  
}
\label{fig:4initialcounts}
\end{figure}

We begin by comparing how varying the initial counts of the wild-type and single-mutants affects the probability of observing at least one double-mutant one year after transplant.
Figure \ref{fig:4initialcounts} depicts these probabilities and draws the expected conclusion that increasing the number of single-mutants over wild-type particles initially present has a larger effect on the occurence of a double-mutant.
This result has clinical relevance as it places a bound on how mutagenic the lentiviral gene addition can be before deleterious effects can be expected.
If the mutagenesis is rare, producing only one to ten single-mutants in the transplanted population, then there is a very small risk that a deleterious second mutation event will occur within a year.
Additionally, the number of transplanted wild-type stem cells can be quite large before mutational events become likely, which bodes well for transplanting large populations of stem cells.

\begin{figure}[ht]\begin{center}
\includegraphics[width = 0.75\textwidth]{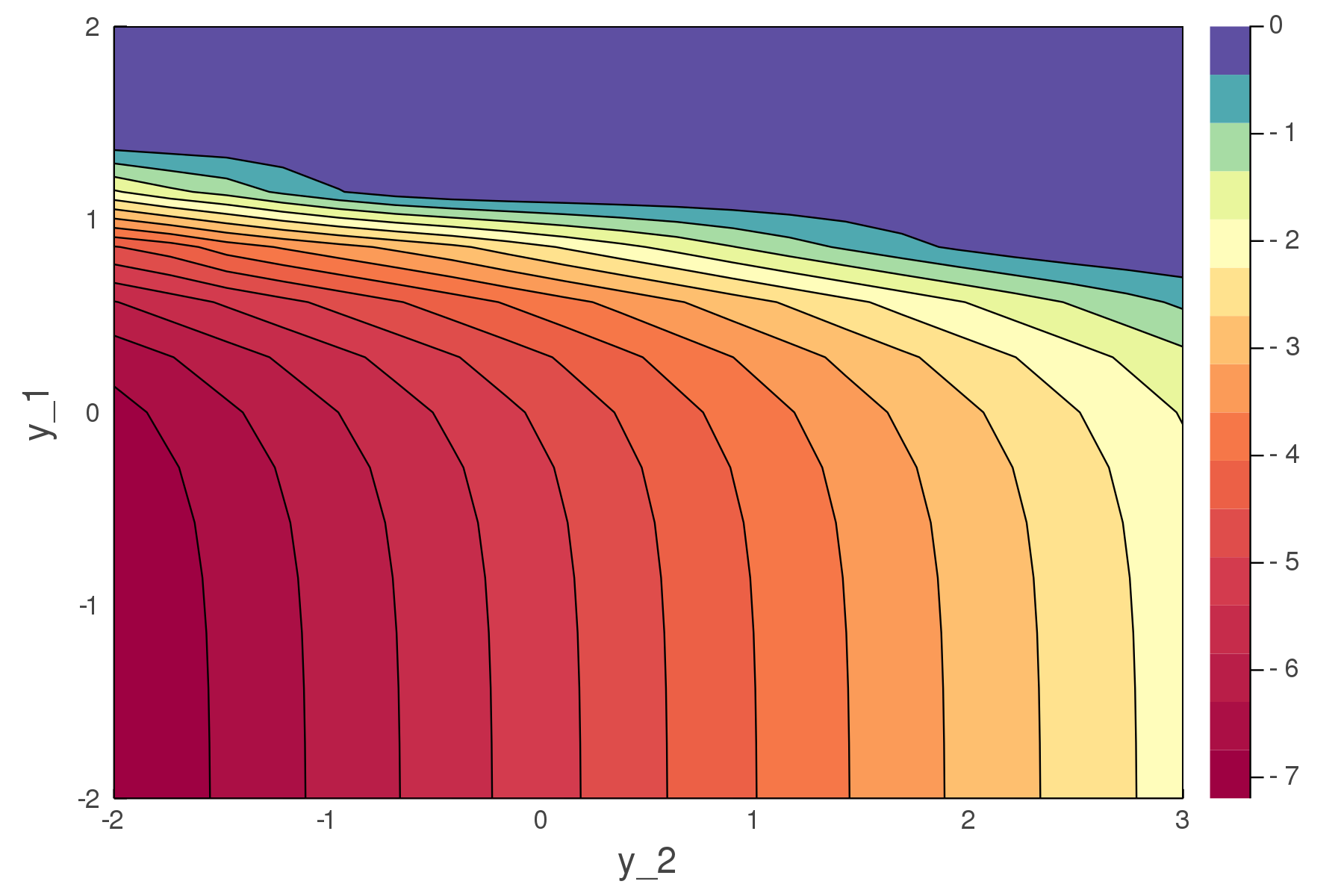}
\end{center}
\caption{Log probability of observing at least one double-mutant at $t = 52$ weeks while varying the birth rates $\lambda_i$ of $X_1$ and $X_2$.  The probability is depicted in log scale.  We have $\lambda_i = 0.024 \times 10^{y_i}$, e.g. when $y_i = 0$, $\lambda_i = 0.024$.  The initial counts are $X_1(0) = 1000$ and $X_2(0) = 0$.}
\label{fig:4lambdas}
\end{figure}

Next we compare varying the birth rates across the population of wild-type stem cells and single-mutants.  
Figure \ref{fig:4lambdas} assumes no initial single-mutants, and so the birth rate of the wild-type population $X_1$ is rate-limiting on the probability of observing a double-mutant.  
We observe that changing the wild-type birth rate by a factor of ten substantially increases the odds of observing a double-mutant, while changing the single-mutant birth rate has a much smaller effect in comparison.  
This finding suggests it is relatively safe for the single-mutations to be quite proliferative if the lentiviral gene addition avoids creating any single-mutants in the original insertion process.

\begin{figure}[ht]\begin{center}
\includegraphics[width = 0.75\textwidth]{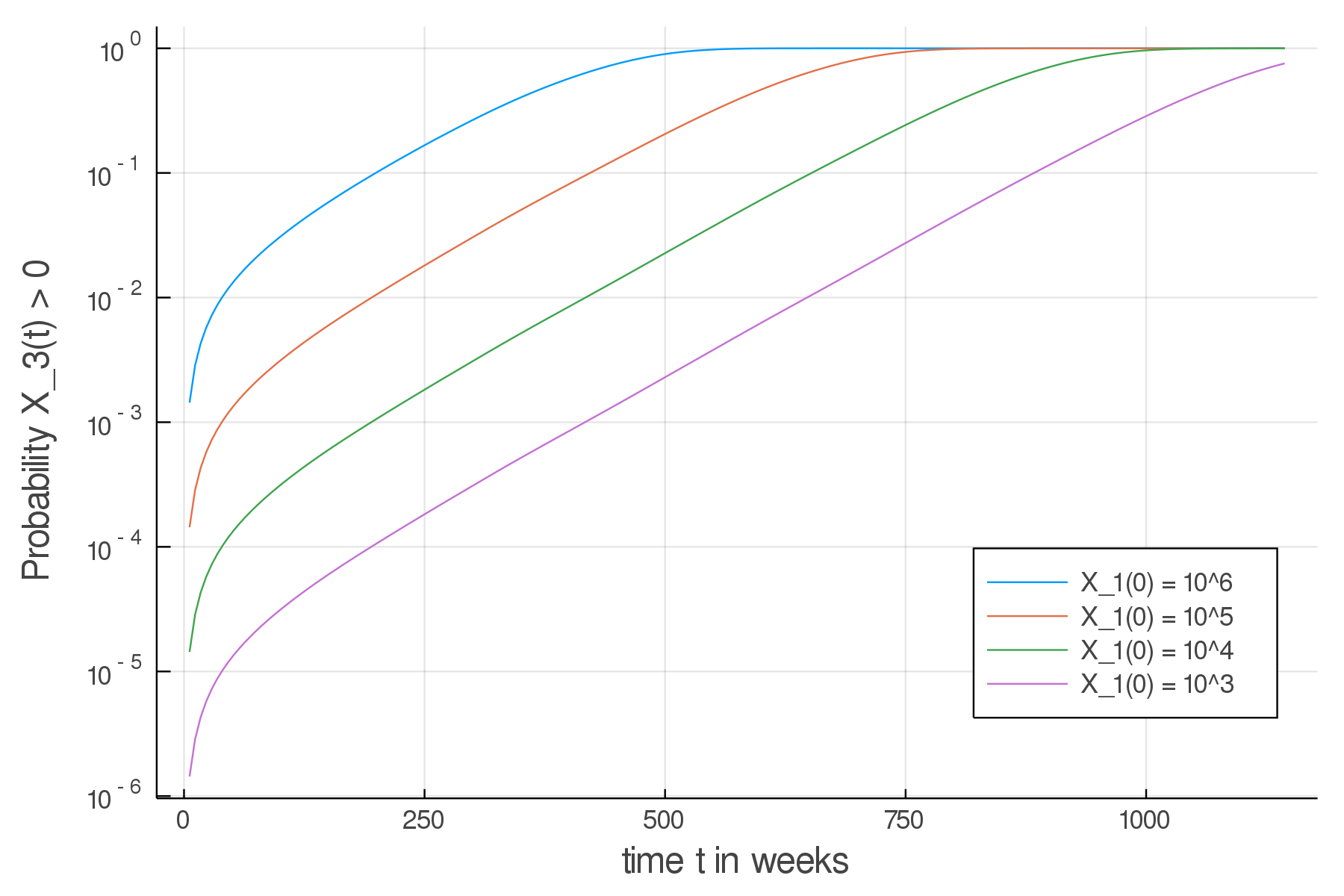}
\end{center}
\caption{Probability of observing at least one double-mutant over time while varying the initial counts of $X_1$ and $X_2$.  The probability is depicted in log scale.  }
\label{fig:4timeseries}
\end{figure}

Since we are solving the ODEs for the univariate PGFs forward in time, we can efficiently examine how the probability of observing a double-mutant changes over time based off of varying initial population counts, without any additional computation. This information is immediately available from the intermediate steps of the ODE solver, visualized in Figure \ref{fig:4timeseries}.  
Qualitatively, we see three phases of growth corresponding to a fast initial phase, followed by an exponential growth in the probability, and finally tapering out towards one as a double-mutant becomes guaranteed to be observed. We also observe that the duration of the second phase depends on the initial population count.

\section{Conclusions}
\label{conc}

This paper proposes ways to numerically evaluate marginal probabilistic quantities based on the PGF $\phi_{ijk}$ of a three-type branching process.
Our technique is exact up to numerical error in inverting the FFT integral, and we show how many clinically relevant questions can be reduced to tractable univariate problems. For instance, in our experiments we are concerned primarily with the double-mutant population $X_3$, and our method yields information about the probabilities of leukemogenesis after gene therapy. 
This presents important information for clinicians about how leukemia may develop under a variety of conditions including mutagenesis from the initial gene insertion, as well as through differing proliferative rates among the mutant and wild-type populations.


Our methodology is general in that we can alter the structure of the underlying mathematical process without significantly changing the numerical processes used to obtain the transition probabilities.
For example, both the shift process $X_1 \rightarrow X_2$ and the mutational birth process $X_1 \rightarrow X_1 + X_2 $ can be incorporated into our system of equations.
These changes will change the structure of the underlying differential equations to be solved, but they do not change the overall numerical manipulations required to solve for and invert the probability generating function. This direct numerical technique provides an attractive alternative to computationally intensive simulation-based approaches. This  computational efficiency becomes critical in studies requiring exploration of many parameters, iterative optimization, or within statistical inferential schemes.

Our approach applies to any multitype branching process model under the standard rate-linearity assumption, enabling us to calculate a wide variety of pertinent transition probabilities, expectations and higher moments toward exploring the transient behavior of the particles. \jx{We see that the framework can be used to answer questions about the population of any given type $X_i(t)$ in a general multitype process. Similarly, one can pose related questions related to sums or differences of populations, such as the probability that one population becomes dominant or outnumbers another \cite[see 13.3.1]{lange2010applied}.  In some cases, we can establish connections to related models and results in the literature. For instance, an anonymous reviewer remarks that we may recognize the equation corresponding to the final type as the birth-death equations, for which additional tools are available. Indeed, if one investigates the probability of a double-mutant in the three stage model depicted in Figure 2, the problem can be equivalently case in terms of survival probabilities of birth-death processes can be used \cite{denes1996exact,crawford2012transition}. } Exploring further connections and possible extensions of these techniques to systems with non-linear interaction dynamics will be a fruitful avenue for future work.   

\section*{Appendix}

\jx{To derive the backward equations, we begin by rearranging \eqref{eq:4phi1}, divide by $t$, then send $t \downarrow 0$ to get an expression for the derivatives of the single-particle PGFs in terms of pseudo-generating functions \eqref{eq:4pseudo},
\begin{equation}\label{eq:4derivs}
 \frac{\dx \phi_1(t, s_1, s_2)}{\dx t} \Big|_{t=0} = u_1(s_1, s_2), \qquad  \frac{\dx \phi_2(t, s_1, s_2)}{\dx t} \Big|_{t=0} = u_2(s_1, s_2)  . 
\end{equation}

Now we may apply the Chapman-Kolmogorov equations for $\phi_1$ along with \eqref{eq:4PGFIndep} to find that 
\begin{align}
\phi_1(t + h, s_1, s_2) & = \sum_{k=0}^\infty \sum_{l=0}^\infty  P_{(1,0),(k,l)}(t + h) s_1^k s_2^l 	\nonumber \\
& = \sum_{k=0}^\infty \sum_{l=0}^\infty \left[ \sum_{i=0}^\infty \sum_{j=0}^\infty P_{(1,0),(i,j)}(t) P_{(i,j),(k,l)}(h) \right]s_1^k s_2^l   \nonumber \\ 
& = \sum_{i=0}^\infty \sum_{j=0}^\infty P_{(1,0),(i,j)}(t) \left[ \sum_{k=0}^\infty \sum_{l=0}^\infty  P_{(i,j),(k,l)}(h) s_1^k s_2^l \right]  \nonumber \\
& = \sum_{i=0}^\infty \sum_{j=0}^\infty P_{(1,0),(i,j)}(t) \phi_{ij}(h, s_1, s_2) \nonumber \\
& = \sum_{i=0}^\infty \sum_{j=0}^\infty P_{(1,0),(i,j)}(t) \phi_1(h, s_1, s_2)^i \phi_2(h, s_1, s_2)^j \nonumber \\
& = \phi_1 \big( t, \phi_1(h, s_1, s_2), \phi_2(h, s_1, s_2) \big) . \nonumber
\end{align}
The relationship is time-symmetric, therefore we also have
\begin{equation} \label{eq:4CKeqn}
\phi_1(t + h, s_1, s_2) = \phi_1 \big( h, \phi_1(t, s_1, s_2), \phi_2(t, s_1, s_2) \big) .
\end{equation}
Naturally the same relationships exist for $\phi_2$.

We now derive the backward Kolmogorov equations for $\phi_1$ and $\phi_2$,  obtained from a Taylor expansion of $\phi_1$ around $t$   
together with \eqref{eq:4derivs} and \eqref{eq:4CKeqn},
\begin{align}
\phi_1(t + h, s_1, s_2) &= \phi_1(t, s_1, s_2) + \frac{\dx \phi_1(t + h, s_1, s_2)}{\dx t} \Big|_{h=0} h + o(h) \nonumber \\
&= \phi_1(t, s_1, s_2) + \frac{\dx \phi_1 \big( h, \phi_1(t, s_1, s_2), \phi_2(t, s_1, s_2) \big)}{\dx t} \Big|_{h=0} h + o(h) \nonumber \\
&= \phi_1(t, s_1, s_2) + u_1 \big( \phi_1(t, s_1, s_2), \phi_2(t, s_1, s_2) \big) h + o(h) \nonumber . 
\end{align}
The same actions produce an analogous expression for $\phi_2$.  
Performing the previous rearrangement, dividing by $h$, and sending $h$ to zero gives us our desired system of backwards Kolmogorov equations.}

%
%


%
%

\bibliographystyle{spmpsci}      
\bibliography{biblio}   

%
%

\end{document}